\documentclass[fleqn,usenatbib]{mnras}
\usepackage{newtxtext,newtxmath}
\usepackage[T1]{fontenc}
\usepackage{ae,aecompl}
\usepackage{graphicx}	
\usepackage{amsmath}	
\usepackage{amssymb}	
\usepackage[normalem]{ulem}

\newcommand{\kms}{\,{\text km s}$^{-1}$}
\newcommand{\rate}{\,{\text Gpc}$^{-3}$\,{\text yr}$^{-1}$}
\newcommand{\Msun}{\,{\text M}$_{\odot } $}

\title[Merger rates in primordial black hole clusters]{Merger rates in primordial black hole clusters without initial binaries}
\author[V.~Korol et al.]{Valeriya Korol$^{1,2,3}$\thanks{E-mail:korol@star.sr.bham.ac.uk}, Ilya Mandel$^{4,5,2,3}$, M.~Coleman Miller$^{6,3}$, Ross P.~Church$^{7,3}$, 
\newauthor and Melvyn B.~Davies$^{7,3}$\\
\\$^1$Leiden Observatory, Leiden University, PO Box 9513, 2300 RA, Leiden, the Netherlands
\\$^2$School of Physics and Astronomy \& Institute for Gravitational Wave Astronomy, University of Birmingham, Edgbaston, \\ 
Birmingham B15 2TT, United Kingdom
\\$^3$Dark Cosmology Centre, Niels Bohr Institute, University of Copenhagen, Juliane Maries Vej 30, DK-2100, K\o benhaven \o, Denmark
\\$^4$School of Physics and Astronomy, Monash University, Clayton, VIC 3800, Australia 
\\$^5$OzGrav: The ARC Centre of Excellence for Gravitational Wave Discovery, Australia 
\\$^6$Department of Astronomy, University of Maryland, College Park, MD 20742-2421, USA
\\$^7$Lund Observatory, Department of Astronomy \& Theoretical Physics, Lund University, Box 43, SE-221 00 Lund, Sweden}
\date{October 21, 2019}

\date{Accepted 2020 June 4; Revised: 2020 May 27; Received: 2019 November 8.}

\pubyear{2020}

\begin{document}
\label{firstpage}
\pagerange{\pageref{firstpage}--\pageref{lastpage}}
\maketitle

\begin{abstract}
Primordial black holes formed through the collapse of cosmological density fluctuations have been hypothesised as contributors to the dark matter content of the Universe.  At the same time, their mergers could contribute to the recently observed population of gravitational-wave sources. We investigate the scenario in which primordial black holes form binaries at late times in the Universe. Specifically, we re-examine the mergers of primordial black holes in small clusters of $\sim 30$ objects in the absence of initial binaries.  Binaries form dynamically through Newtonian gravitational interactions.  These binaries act as heat sources for the cluster, increasing the cluster's velocity dispersion, which inhibits direct mergers through gravitational-wave two-body captures.  Meanwhile, three-body encounters of tight binaries are too rare to tighten binaries sufficiently to allow them to merge through gravitational-wave emission.  We conclude that in the absence of initial binaries, merger rates of primordial black holes in the Bird et al. (2016) initial cluster configuration are at least an order of magnitude lower than previously suggested, which makes gravitational-wave detections of such sources improbable.  

\end{abstract}




\section{Introduction}

The concept of black holes formed directly from the gravitational collapse of the cosmological density fluctuations in the early Universe, known as primordial black holes (PBHs), dates back to the 1970s \citep{Hawking71}. If the density fluctuations collapsed before $\sim$1$\,$s after the Big Bang, the baryons that produced them would be locked up and thus there would not be a conflict with light element nucleosynthesis.
Then PBHs would behave as non-baryonic cold dark matter (DM) throughout the subsequent evolution of the Universe. 
The large variety of mechanisms that could have produced the density fluctuations that seed PBHs yield possible mass functions that extend from the Planck mass to the mass of a galaxy cluster \citep[see][for reviews]{Carr2016,car20}. 
In principle, all of the DM in the Universe could be PBHs.
The abundance of PBHs in different mass regimes have been strongly contested in recent years by a number of astrophysical and cosmological experiments, leaving only three mass windows in which PBHs could still provide an important contribution to the DM: asteroid mass PBHs ($10^{16} - 10^{17}\,$g), sub-lunar mass PBHs ($10^{20} - 10^{26}\,$g) and stellar mass PBHs (20 - 100\Msun) \citep{car20}.

Recently, detections of gravitational waves (GWs) from merging black holes of $10 - 40$\Msun\ by  Advanced LIGO (Laser Interferometer Gravitational-wave Observatory) and Virgo have returned attention to stellar mass PBHs \citep{ligo2016,ligo2018}.
Shortly after the first LIGO detection (GW150914) several authors suggested that it might have been a merger of PBHs.   There are a number of ways to produce PBH binaries. These can be divided roughly into two channels: initial binaries produced in the early Universe \citep[e.g.,][]{sas16,Ali2017} and binaries formed much later through dynamical interactions \citep[e.g.,][]{Bird2016,Clesse2016}. 
In the initial binaries channel, PBHs can form in binaries as a result of chance proximity, as first described in \citet{nak97}. 
In the second channel, single PBHs that formed in the early Universe form binaries much later due to dynamical interactions of PBHs in DM halos.
We note that constraints on PBH abundance from gravitational wave observations are based mostly on the early Universe scenario, because later production of PBH binaries is highly subdominant \citep[e.g.][]{Ali2017}. 
However, because the early Universe channel has a number of caveats and uncertainties, in this paper we revisit the dynamical interaction channel.

Let us assume that $30\,$M$_\odot$ PBHs form clusters, and that PBH clusters have the large scale structure distribution of DM halos (e.g., the \citet{pre74} mass function).
For each PBH cluster there is a finite probability that two initially unbound PBHs pass close enough to gravitationally capture each other and form a bound binary system through the emission of a burst of gravitational waves at periapsis.
Following this idea \citet{Bird2016} derived a PBH capture rate between $10^{-4}$ and $1400$\rate\ depending on assumptions about how PBHs cluster in DM halos. 
This broad range encompasses the best estimate based on all the events of the first and the second advanced detector observing runs, $53.2^{+58.2}_{-28.8}$\rate \citep{ligo2018}.
In particular, \citet{Bird2016} find that the major contribution to the total merger rate of PBH binaries comes from clusters of $\sim10^3$\,M$_\odot$, while smaller DM clusters have too few PBHs and evaporate promptly due to weak gravitational interactions. 
Gravitational two-body captures are more efficient in low-mass PBH clusters, because PBHs in low-mass clusters move more slowly and because small clusters are more concentrated. While the first statement is based on dynamical considerations, the second statement is a consequence of the hierarchical formation of DM halos in the $\Lambda$CDM cosmological model: because low-mass halos assemble earlier, when the mean density of the Universe is higher, they have higher concentrations than high-mass halos \citep[e.g.,][]{nfw97,wec02}. 

\citet[][see also earlier work of \citealt{afsh03}]{Ali2017} pointed out that due to the discrete nature of PBHs, in clusters composed of $\lesssim1000$ PBHs, Poisson perturbations dominate over standard adiabatic perturbations \citep[considered in][]{Bird2016}. 
This implies that low-mass PBH clusters form much earlier, and, therefore, they are also denser. As a consequence, using scalings appropriate for large-$N$ dynamical systems, \citet{Ali2017} conclude that clusters with $\lesssim 3 \times 10^4$\,M$_\odot$ also evaporate by the present time and do not contribute to the merger rate of PBHs. 

In low-mass PBH clusters, binaries can also form through non-dissipative three-body interactions, in which one PBH removes enough kinetic energy to leave the other two in a bound state.
Once formed, binaries serve as a source of energy in a cluster \citep{heg93}. By interacting with single PBHs (and other binaries, if present) they heat the cluster, which then expands and partially evaporates (i.e., loses objects).  This enhances the rate of evaporation relative to that obtained by only considering weak gravitational interactions, as in \citet{Bird2016}.   
Moreover, cluster expansion causes the density of the cluster and the merger rate through two-body capture to decrease (see Eq.\eqref{eqn:2brate}).   

In this work we perform a suite of N-body simulations to quantify the effect of dynamical interactions on the PBH merger rate.  We focus on low-mass, low-$N$ clusters for which analytical scalings (see \citealt{Ali2017} and section \ref{sec:theory} below) may break down.
Specifically, we investigate what binary formation process provides the major contribution to the PBH merger rate in few $\times 10^2$ - $10^3\,$M$_\odot$ clusters. 
We find that in this mass regime the first hard binary is formed within a few hundred Myr.  
After the formation of the first binary the cluster expands by a factor of 20 in a time comparable to the age of the Universe.
Consequently, the rate of two-body captures at the present time drops by an order of magnitude.

Spatial clustering of PBHs has been discussed in light of current LIGO observations \citep[e.g.,][]{rai17,bal18,bringmann18,inm19}. In this paper we investigate the dynamics for the specific initial PBH distributions proposed by \cite{Bird2016}.  
We refer the reader to \citet{bringmann18} for a discussion of plausible PBH distributions.

This paper is organised as follows.
In Section~\ref{sec:theory} we present theoretical considerations using as an example a $450\,$M$_\odot$ cluster composed of $15 \times 30\,$M$_\odot$ PBHs.
In Section~\ref{sec:simulations} and \ref{sec:results} we describe the setup for N-body simulations and present our results.
In Section~\ref{sec:discuss} we discuss the implications of these results for the merger rate of PBHs, and present our conclusions.


\section{Theoretical considerations} 
\label{sec:theory}

In this section we consider an example of an N-body non-relativistic interacting system composed of $N=15\,$PBHs, which we call a {\it cluster}. For simplicity, we assume that the cluster is composed of a single mass-species of $m=30\,$M$_{\odot}$, such that the total mass of the cluster is $M=Nm = 450\,$M$_{\odot}$.
This example falls in the mass-range that provides the major contribution to the PBH merger rate in \citet{Bird2016}\footnote{Clusters with $M<400\,$M$_\odot$ are expected to evaporate in a few Gyr after their formation, and thus are not included in the merger rate derivation.}.
Finally, we assume the cluster to be spherically symmetric with radius $R=2\times 10^5\,$au (i.e. 1\,pc).

The number density of PBHs in the cluster is 
\begin{equation}
    n \sim \frac{N}{R^3} = 15\,{\rm pc}^{-3} \approx 2 \times 10^{-15}\,{\rm au}^{-3}
\end{equation}
and the typical velocity dispersion  is 
\begin{equation}
    v_{\rm disp} \sim \sqrt{\frac{GM}{R}} \approx 1.5\,{\rm km~s}^{-1} \approx 0.3\,{\rm au\,  yr}^{-1}.
\end{equation}
Next, we consider a binary system inside the cluster with orbital separation (semi-major axis) $a$ and orbital speed $v_{\rm orb} \sim \sqrt{Gm/a}$. If the binding energy of the binary is smaller in magnitude than the typical kinetic energy of PBHs in the cluster, the binary is called {\it soft}. Soft binaries are likely to be disrupted by interactions with single PBHs. In the opposite case the binary is called {\it hard}, and dynamical interactions with single PBHs will further tighten the binary.
Thus, hard binaries in a cluster typically survive encounters and tend to become harder, whereas soft binaries tend to split \citep{heggie75,hil75}. The hard/soft boundary depends on the properties of the cluster and can be estimated as 
\begin{equation}
    a_{\rm HS} \sim \frac{R}{N} = 10^4{\rm au}
\end{equation}
in our case.  Even in the absence of primordial binaries, binaries will generally form throughout three-body interactions over the lifetime of the cluster. This happens on the timescale 
\begin{equation}
    \tau_{\rm 3B} \sim \frac{N^2R^{3/2}}{(GNm)^{1/2}} \sim N^2 \tau_{\rm cross} \simeq 150\, {\rm Myr},
\end{equation}
where $\tau_{\rm cross} = R/v_{\rm disp} = 0.7\,$Myr is the cluster crossing timescale.

Alternatively, binaries can form via gravitational two-body captures.
This can happen when two PBHs pass close enough to each other to emit GW radiation.
If the energy released in GWs during the passage exceeds the total initial kinetic energy, the two PBHs become bound. 
The cross section of two-body captures is given by \citep{qui89} 
\begin{equation} \label{eqn:2brate}
 \sigma_{\rm 2B} = 5 \times 10^{-9} \left( \frac{m}{30{\text \Msun}}\right)^2 \left( \frac{v_{\rm disp}}{1.5 \text{  km s}^{-1}}\right)^{-18/7} \quad \text{pc}^2, 
\end{equation}
while the typical timescale is $\tau_{\rm 2B}= (n\sigma_{\rm 2B} v_{\rm disp})^{-1} = 90 \times 10^3\,$Gyr $ \gg \tau_{3B}$. 
Therefore, in the case we consider, three-body interactions will be the dominant binary formation mechanism.

Once formed, hard binaries necessarily interact with single PBHs in the cluster. These 2+1 interactions happen on a timescale of
\begin{equation} \label{eq:tau2+1}
    \tau_{\rm 2+1} \sim \frac{v_{\rm disp}}{nGma},
\end{equation}
so the initial timescale for strong interactions is $\tau_{\rm 2+1} (a_{\rm HS}) \approx 10$ Myr. 
Each interaction carries away a significant fraction of the binary orbital energy and the interloper PBH is ejected with a speed $\sim v_{\rm orb}$. Because in this example the cluster is composed of a single mass species, conservation of linear momentum of the binary - interloper system implies that the binary must be ejected with a speed $\sim v_{\rm orb}/2$. This speed needs to be compared to the escape speed from a cluster, which is typically a few times the velocity dispersion of the cluster \citep[e.g.,][]{bin08}. 
This implies that the recoil kicks will eject the binary once its orbital speed reaches $\sim 10 v_{\rm disp}$.
Since $v_{\rm orb} \sim v_{\rm disp}$ at the hard-soft boundary, and $v_{\rm orb} \propto a^{-1/2}$, the minimum semi-major axis at the ejection $a_{\rm eject}$ is approximately two orders of magnitude smaller than $a_{\rm HS}$ before being ejected. At that time, the 2+1 interaction timescale is  $\tau_{\rm 2+1} (a_{\rm eject}) \sim 1$ Gyr.
Binaries tighter than $\sim 0.01 a_{\rm HS}$ can only remain in the cluster  if gravitational wave hardening takes over as the dominant forcing mechanism before the binary reaches $a_{\rm eject}$ and can be ejected. 
The timescale of gravitational wave hardening a circular binary composed of $30+30\,$M$_\odot$ PBHs is \citep{Peters1964}:
\begin{equation}\label{tgw0}
\tau_{\rm GW}(a,e=0) \simeq 6 \times 10^{20}\, {\rm yr} \left( \frac{a}{100\,{\rm au}} \right)^4.
\end{equation}
If the binary is highly eccentric, with eccentricity $e \to 1$, $\tau_{\rm GW}$ becomes 
\begin{equation} \label{tgw}
\tau_{\rm GW} (a,e) \simeq\frac{768}{425}(1-e^2)^{7/2}\  \tau_{\rm GW}(a,e=0).
\end{equation}

The ultimate fate of the binary is essentially determined by a comparison between $\tau_{2+1}$, the Hubble time $\tau_{\rm H}$, and $\tau_{\rm GW}$:
\begin{itemize}
    \item if $\tau_{\rm 2+1}+\tau_{\rm GW}(e=0) < \tau_{\rm H}$ for some $a \in (a_{\rm HS}, a_{\rm eject})$, the binary will merge inside the cluster through a sequence of 2+1 interactions and GW emission;
    
    \item if $\tau_{\rm 2+1}+\tau_{\rm GW}(e=0)  > \tau_{\rm H}$ but $\tau_{\rm 2+1} < \tau_{\rm H}$ at $a_{\rm eject}$, the binary may either merge inside the cluster if 2+1 interactions happen to drive it to a sufficiently high eccentricity to reduce $\tau_{\rm GW}$ at ejection or may be ejected;
    
    \item if both $\tau_{\rm 2+1} > \tau_{\rm H}$ and $\tau_{\rm GW} > \tau_{\rm H}$ at some $a \in (a_{\rm HS}, a_{\rm eject})$, the binary will remain in the cluster and stall at the orbital separation at which $\tau_{\rm 2+1} > \tau_{\rm H}$.
\end{itemize}

The $N \sim 30$ regime pushes the range of validity of analytical scalings; fortunately, this regime is readily amenable to numerical simulations, which we introduce in the next section.


\section{n-body Simulations} \label{sec:simulations}

In this Section we describe a suite of simulations of clusters with total mass $10^2 - 10^3\,$M$_{\odot}$ performed using {\sc REBOUND}, an N-body open source code \citep{rebound}. 
Specifically, we model three types of clusters:
\begin{enumerate}
\item $15 \times 30\,$M$_{\odot}$ PBHs,
\item $5 \times 30 + 30 \times 10\,$M$_{\odot}$ PBHs,
\item $35 \times 30\,$M$_{\odot}$ PBHs,
\end{enumerate} 
such that types (i) and (ii) have the same total mass of 450\,M$_\odot$, while (ii) and (iii) both consist of 35 PBHs.
We set the size of all clusters to be $1\,$pc.
Note that our clusters are more compact than the reference example of a $450\,$M$_\odot$ DM halo with a velocity dispersion of $0.15\,$\kms \citep{Bird2016}, which implies that the virial radius of the  reference example is $\sim 100\,$pc. 
This changes the approximate numerical values of the scalings derived in the previous section to: $a_{\rm HS} \sim 10^6$ au; $a_{\rm eject} \sim 10^4$ au; $\tau_{\rm 3B} \sim 150$ Gyr $\ll \tau_{\rm 2B} \sim 10^9$ Gyr; $\tau_{2+1} (a_{\rm HS}) \sim 10$ Gyr; and $\tau_{2+1} (a_{\rm eject}) \sim 1000$ Gyr.
Our simulations are essentially scale-free; other than the two-body capture, which has an additional length scale set by the gravitational radius of the PBH masses, the results can be re-scaled to an arbitrary cluster size (for re-scaling to the reference example from \citet{Bird2016}, this corresponds to multiplying lengths by $10^2$, dividing speeds by $10$ and multiplying time scales by $10^3$).  In the following we refer to all quantities as originally set in our simulations as ``simulated'', and we also report ``re-scaled'' quantities to compare with the reference example. 

In the simulations we consider three values of the initial virial ratio of the cluster, defined as
\begin{equation}
    V =  \frac{\sum_{i} m_i v_i^2}{2\sum_{ij} Gm_i m_j/r_{ij} } 
\end{equation}
with $r_{ij}$ being the separation between PBHs $i$ and $j$.
We set $V$ to 0.5, 0.3 and 0.1, allowing us to model clusters with a range of initial properties; however, we find that clusters promptly virialise to $V=0.5$.

In total we performed 90 simulations: 10 for each combination of the three types of cluster (i, ii and iii) and the three initial virial ratios (0.5, 0.3 and 0.1). Our simulations are purely Newtonian; we do not consider additional forces or effects such as general relativity or tidal forces.
We draw the initial positions of PBHs from a uniform space-density distribution and velocities from a Maxwellian distribution with the scale parameter equal to the velocity dispersion of the cluster $v_{\rm disp}$.
We evolve clusters for $100\,$Myr (100 Gyr of re-scaled time) using the {\sc IAS15} integrator \citep{ias5}. 
We record the formation of PBH binaries and multi-body bound systems and their properties.


\section{Results} \label{sec:results}

\begin{figure}
        \centering
	 \includegraphics[width=0.49\textwidth]{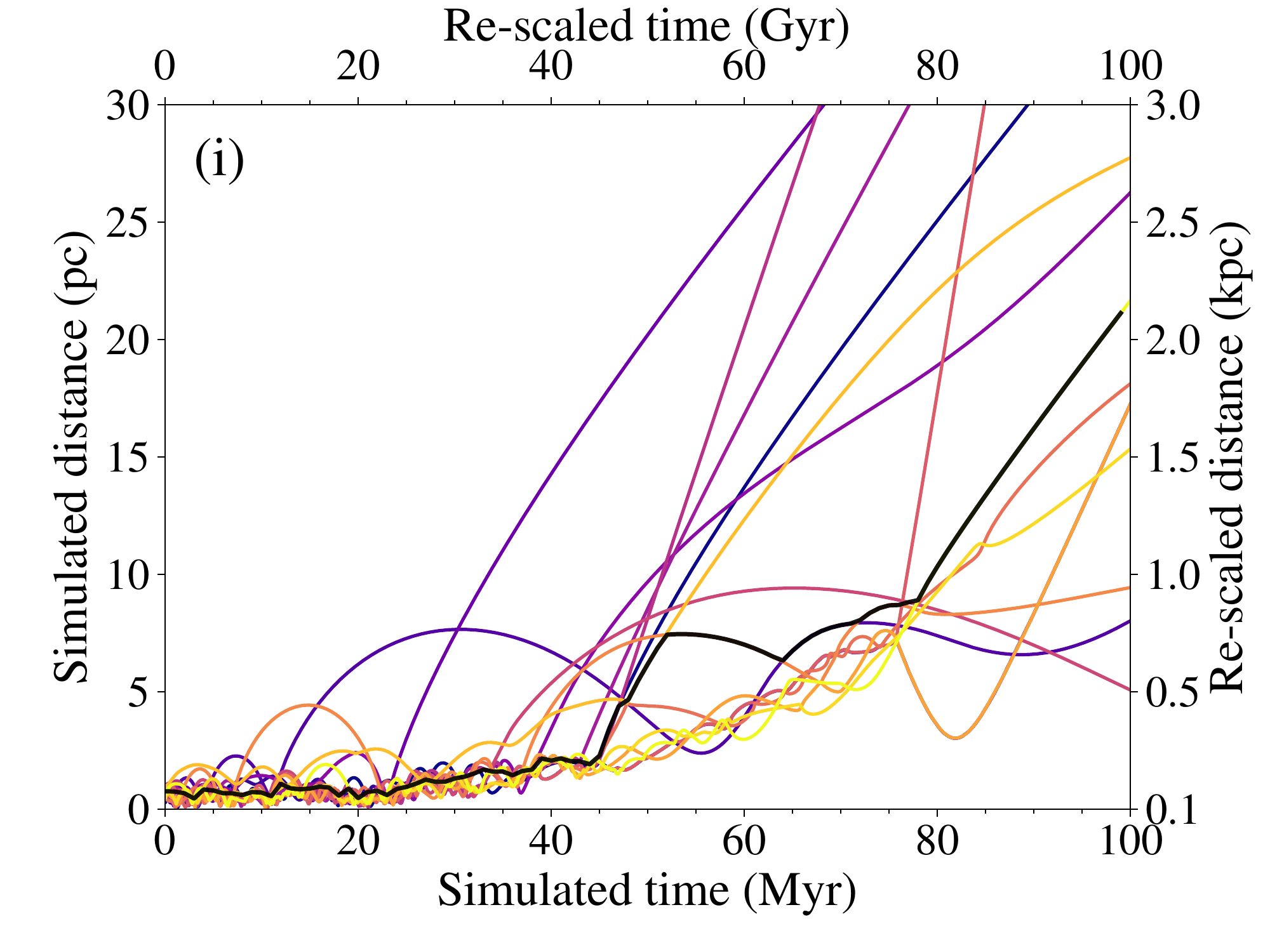} 
	 \includegraphics[width=0.49\textwidth]{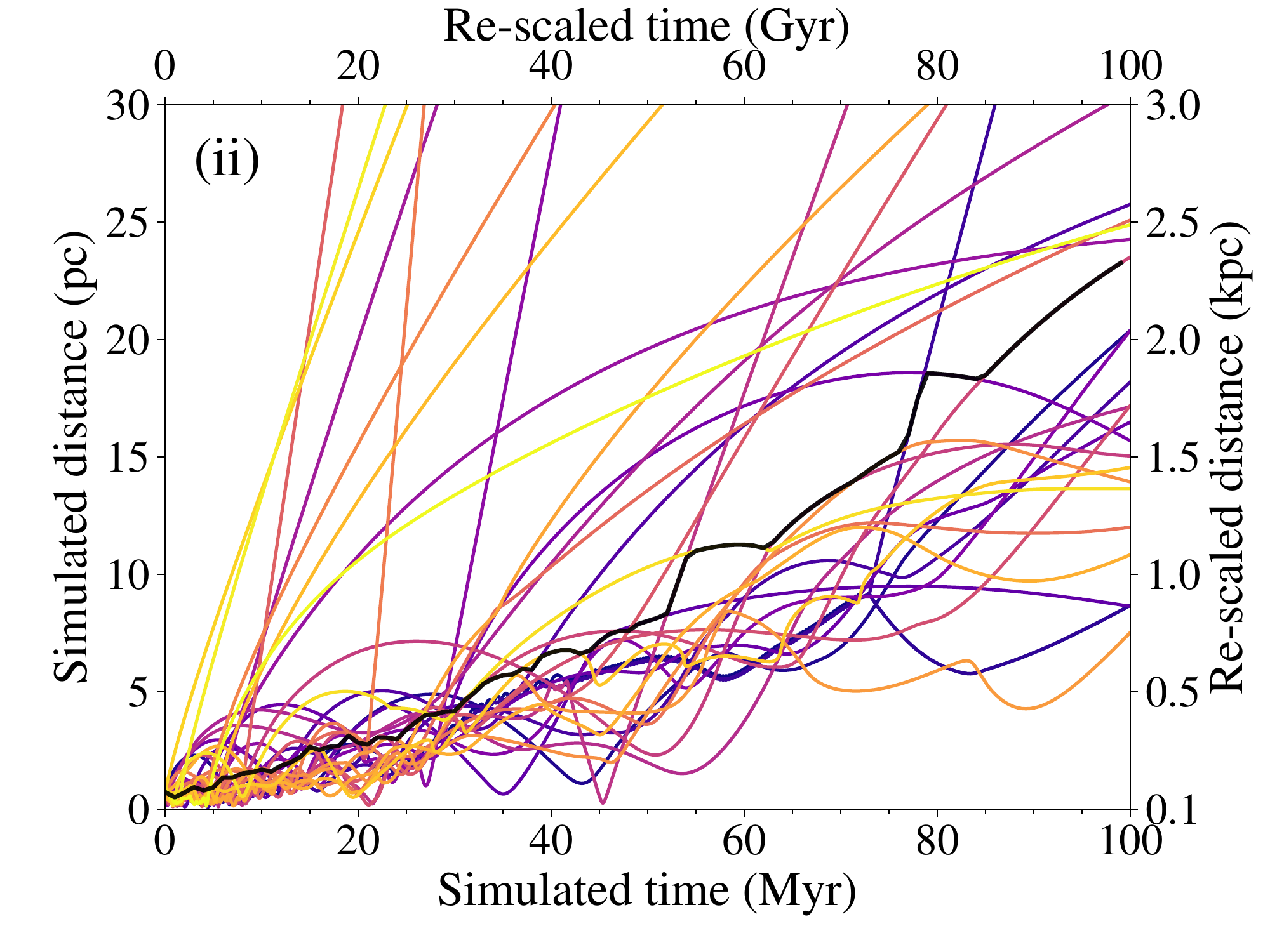}
	 \includegraphics[width=0.49\textwidth]{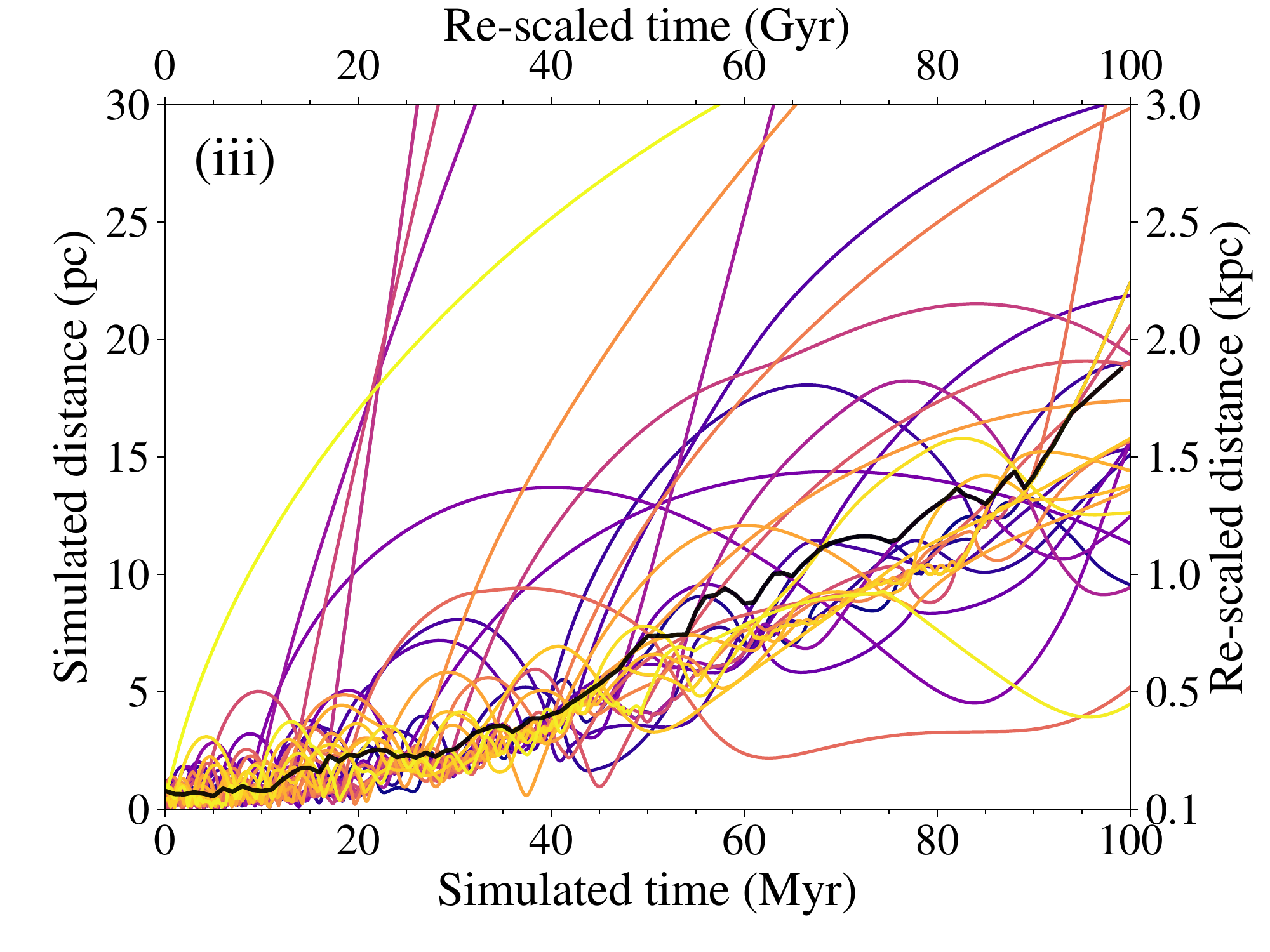}
         \caption{Distance from the barycenter of the cluster as a function of time: each coloured line represents one PBH. The thick black line represents the median distance.
         The bottom x-axis and left y-axis represent our fiducial simulation; the top x-axis and right y-axis correspond to clusters re-scaled to the reference example from \citet{Bird2016}. From the top to the bottom we show the results for clusters with i) $15 \times 30\,$\Msun, ii) $5 \times 30 + 30 \times 10\,$\Msun\, and iii) $35 \times 30\,$\Msun\, PBHs and virial ratio $V=0.5$.}
       \label{fig:1}
\end{figure}

After $100\,$Myr of simulation time we find that the PBH clusters are significantly spread out. Specifically, we find that the median distance of a PBH from the centre of mass of the cluster is more than an order of magnitude larger than it was initially. 
We illustrate an example for each type of cluster with $V=0.5$ in Fig.~\ref{fig:1}. 
Each line represents the distance of PBHs from the cluster's barycenter as a function of time (each coloured line represents a single PBH): from top to bottom for the cluster of a type (i), (ii) and (iii).
The thick black line, representing the median distance of PBHs, clearly shows that clusters expand.
The majority of the (coloured) lines closely intertwine for the first part of the simulation indicating that the cluster stays bound.
Arched lines represent objects that are kicked out of the cluster with speed lower than the escape speed, such that they fall back into the cluster after reaching a maximum distance. 
Nearly vertical lines indicate PBHs ejected from the cluster. 
Note that we do not remove ejected PBHs from the simulation, so the cluster's barycenter drifts as the cluster evolves.

\begin{figure}
        \centering
	     \includegraphics[width=0.49\textwidth]{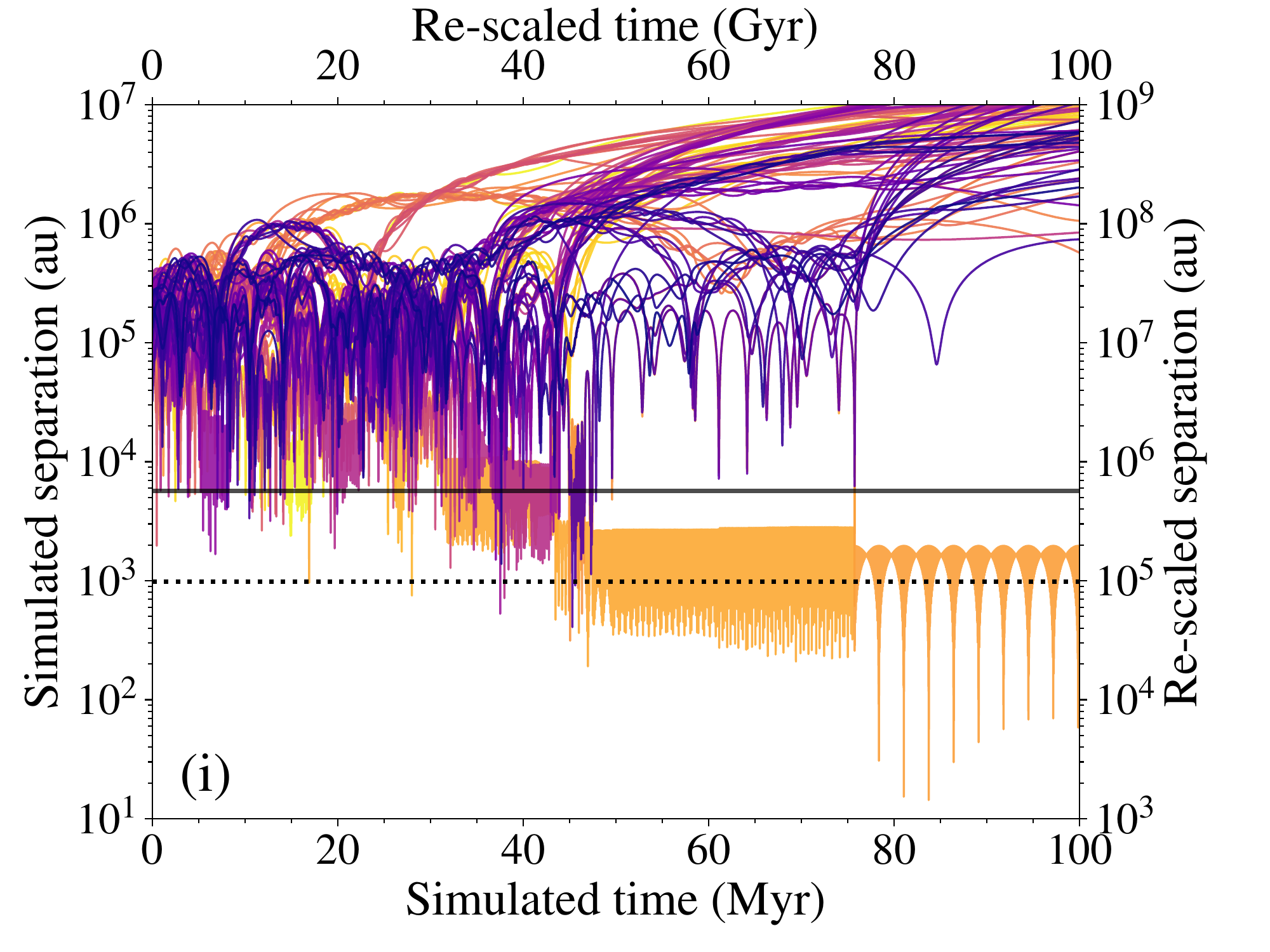} 
	     \includegraphics[width=0.49\textwidth]{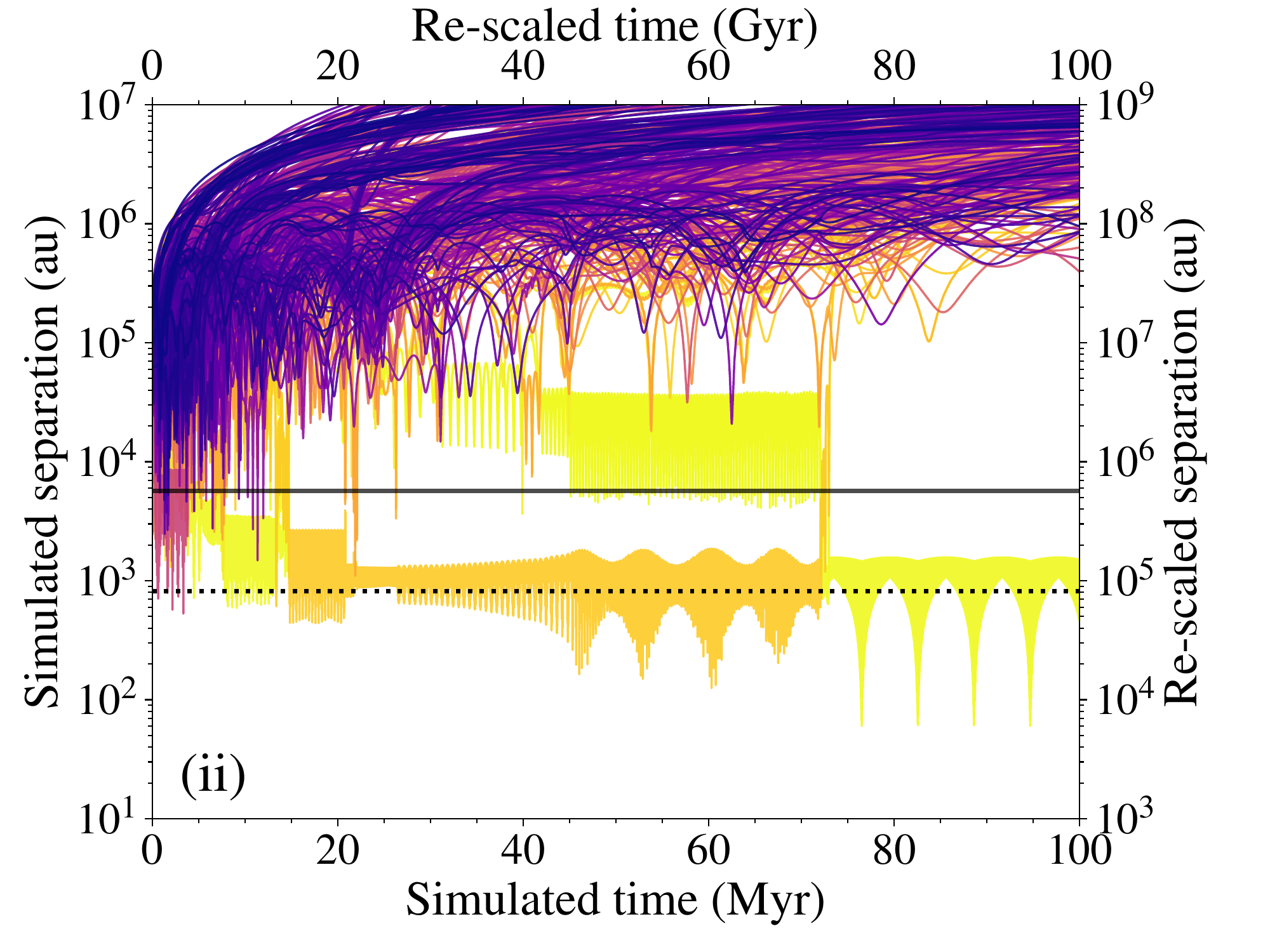}
	     \includegraphics[width=0.49\textwidth]{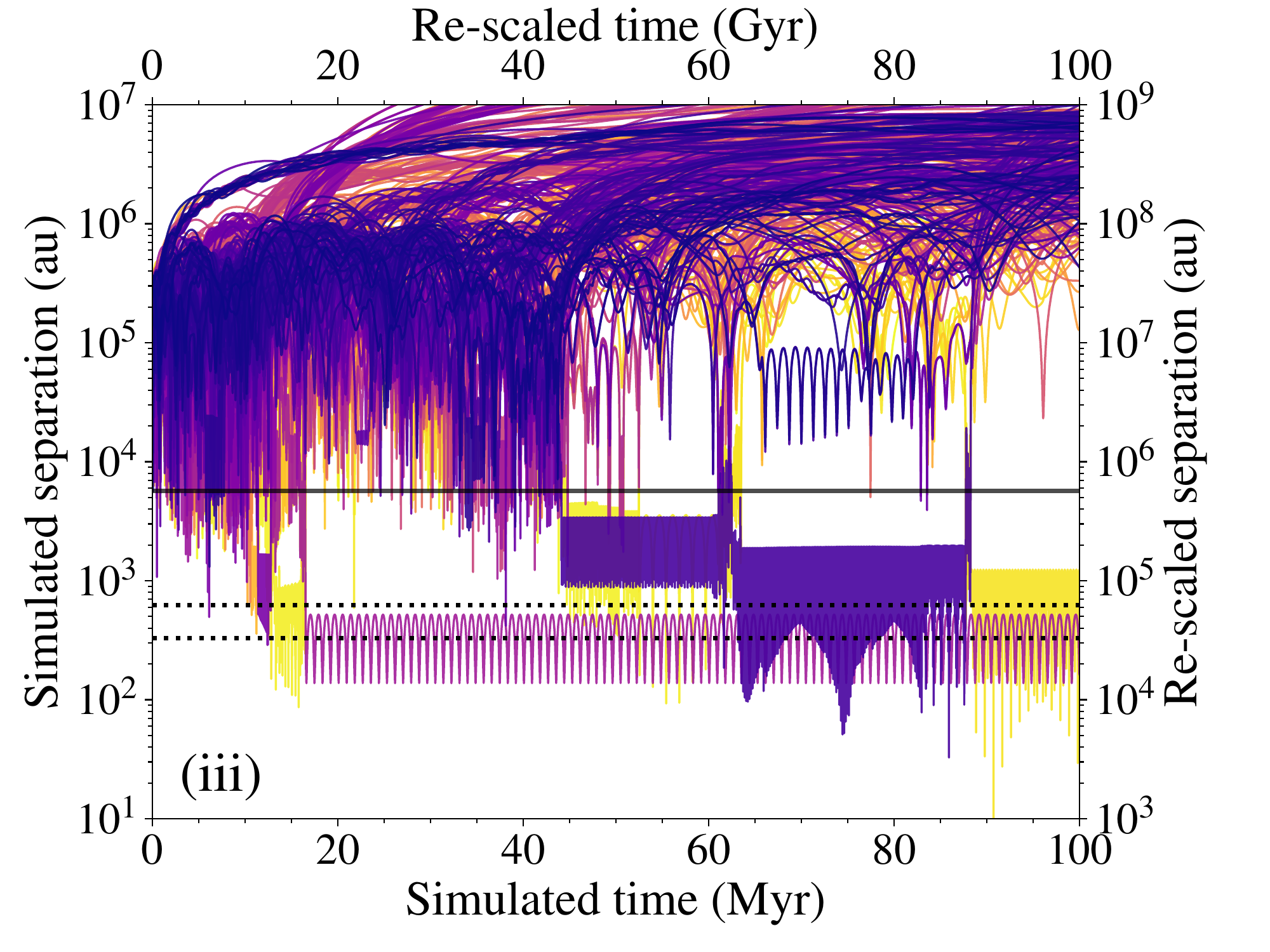}
         \caption{Separation between two PBHs as a function of time: each line represents a PBH pair. The clusters are the same as in Fig~\ref{fig:1}. The horizontal black solid line shows the separation between hard and soft binaries $a_{\rm HS}$. Dotted horizontal lines indicate the semi-major axes of hard binaries formed in our simulations. These three examples illustrate that the typical outcome in our simulations is the formation of a few hard binaries.} 
       \label{fig:2}
\end{figure}

\begin{figure}
\centering
\includegraphics[width=0.5\textwidth]{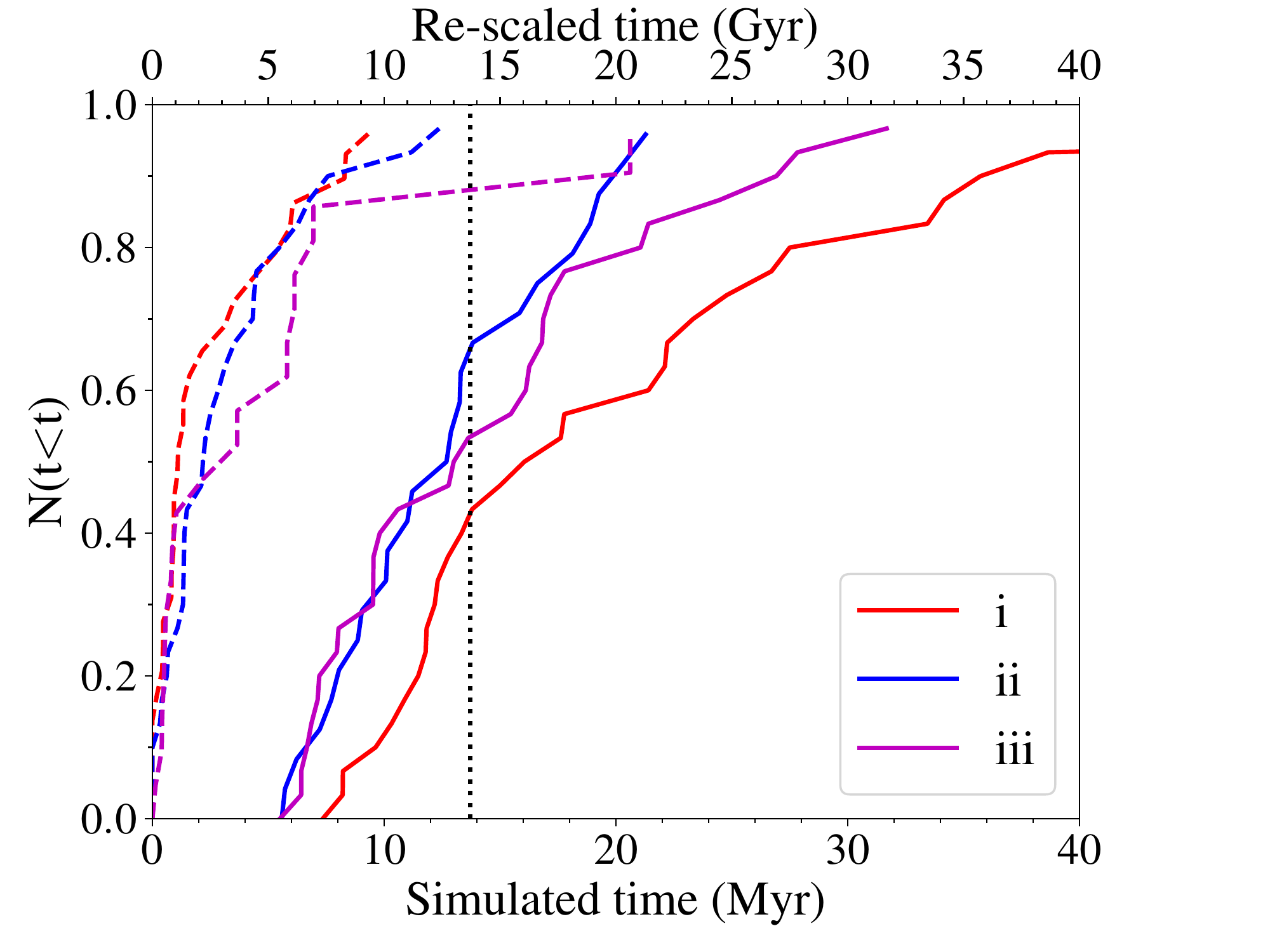} 
\caption{Cumulative distribution of the time when the density of the cluster drops by a factor of 20 (solid lines), which implies a drop in the merger rate due to two-body captures by an order of magnitude, and of the formation time of the first hard binary (dashed lines). 
The vertical dotted line marks $t=13.7\,$Gyr for the re-scaled simulation.}
\label{fig:3}
\end{figure}

\begin{figure*}
\centering
\includegraphics[width=0.75\textwidth]{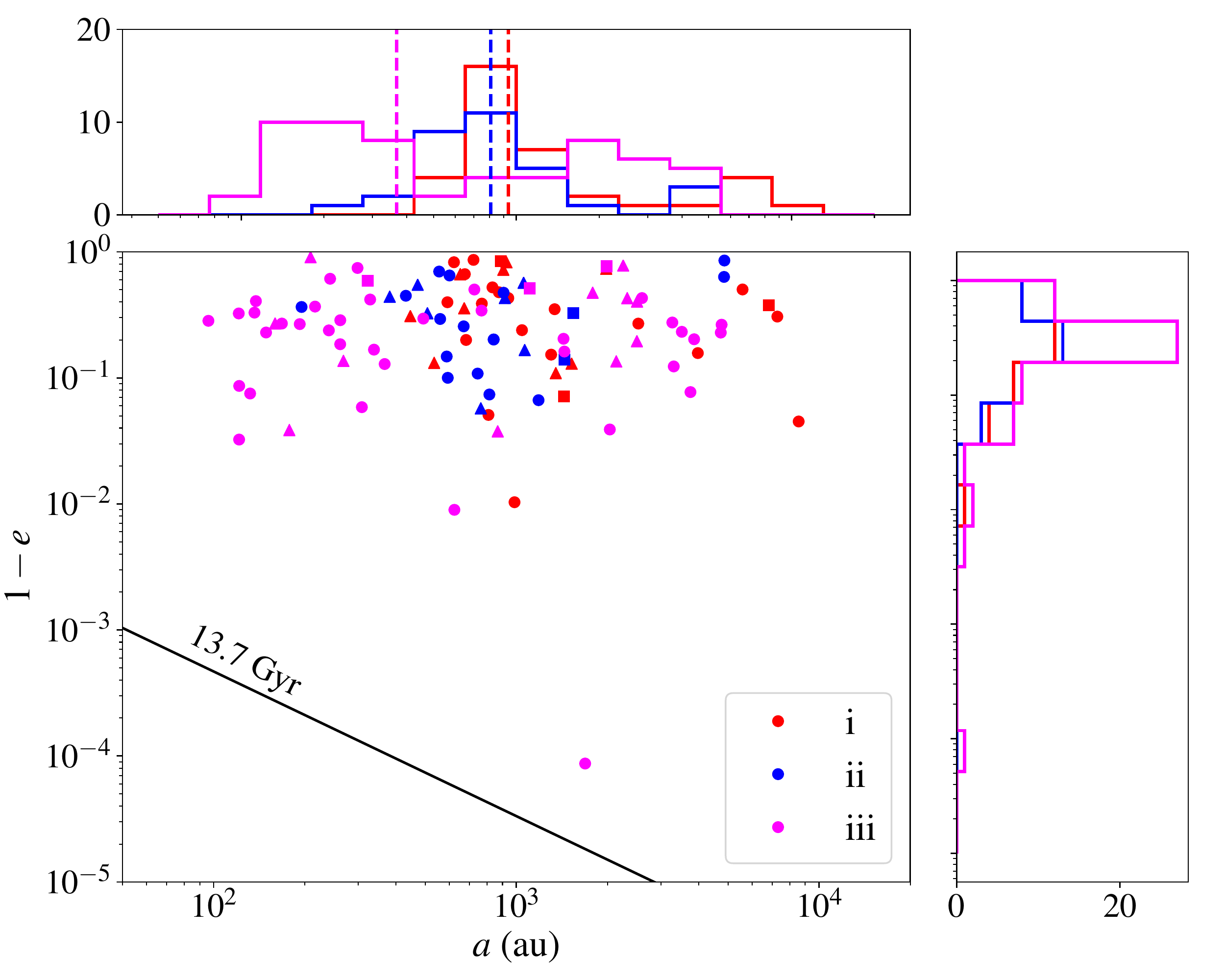} 
\caption{Properties of the binaries formed in our simulations: circles are binaries that are not bound to any other PHB, triangles are triples and squares are multiple systems. Red, blue, and magenta denote simulations of clusters of types (i), (ii), and (iii), respectively. The black solid line shows the merger time of $13.7\,$Gyr for a $30 + 30\,$M$\odot$ PBH binary. In the top and the right panels we show respectively distributions of semi-major axis and eccentricities. Dashed vertical lines indicate median values of $a$. To re-scale to the \citet{Bird2016} cluster size, the abscissa must be scaled by a factor of 100.} 
\label{fig:4}
\end{figure*}

In Fig.~\ref{fig:2} we plot the separation between each pair of PBHs in the cluster, i.e. $|{\bf r}_{\rm i} - {\bf r}_{\rm j}|$ where ${\bf r}$ is the position vector of the object in the cluster's barycenter reference frame.
The black solid horizontal line represents the hard/soft boundary, $a_{\rm HS}$. Consequently, coloured lines that lie below represent hard PBH binaries.  
Dotted horizontal lines indicate the semi-major axis of hard binaries at $t=100\,$Myr.
These three examples illustrate that the typical outcome in our simulations is the formation of a few hard binaries.
We also find that in clusters of type (ii), composed of a mix of PBHs of mass 10$\,$M$_{\odot}$ and 30$\,$M$_{\odot}$, the more massive PBHs tend to form binaries, consistent with expectations for mass segregation and substitutions during $2+1$ interactions.

Binaries represent an energy source in a cluster, and dynamical interactions with binaries cause the cluster to heat up, expand and in some cases evaporate \citep{heg93}.
For example, the binding energy of the hard binary in the top panel of Fig.~\ref{fig:2} with a separation of $\sim 0.1\, a_{\rm HS}$ is $\sim67$\% of the binding energy of the entire cluster. 
This means that to harden this binary through 2+1 interactions, which transfer energy from the binary to other cluster members, the cluster expanded by a factor of about 3.
Thus, the density and typical speed of the PBHs in the cluster decrease respectively by $1/3^3$ and  $1/\sqrt{3}$.
Therefore, the merger rate by gravitational two-body captures drops by more than an order of magnitude ($\Gamma_{\rm 2B} \propto n^{31/42}$, see Section~\ref{sec:theory}).
If the hard binary is ejected without having time to share its energy with the cluster, the expansion rate of the cluster would be more moderate than in this particular example.

In order to suppress the merger rate due to two-body captures by an order of magnitude the density needs to drop by a factor of 20.
We use this scaling to estimate how long it takes to suppress the two-body capture merger rate by an order of magnitude in our simulations. Specifically, we trace the evolution of the density of PBHs in the cluster with time. 
We use the density within half-mass radius, the radius measured from the cluster's barycentre containing half the total mass of the cluster.
In Fig.~\ref{fig:3} we plot the cumulative distributions of the time when $\rho(t)/\rho(t=0) = 1/20 $ (solid lines) and of the formation time of the first hard binary (dashed lines) across simulations.
We find that $\gtrsim 90\%$ of clusters form the first hard binary within $10\,$Myr. 
Thereafter the hard binary heats the cluster and causes the density to decrease by a factor of 20 within $40\,$Myr.  Thus, all simulated clusters evaporate in far less than a Hubble time.

By re-scaling these results to the reference example we find that the formation of the first hard binary occurs within 10\,Gyr, and that by $t=13.7\,$Gyr (dashed vertical line in Fig.~\ref{fig:3}) the cluster density and hence the rate of mergers through two-body captures drops significantly in 40 - 60\% of the clusters.  
We find that the density drops by a factor between 2 and 1100, with a median of 20. 
This corresponds to a drop in the merger rate of about 10.

\cite{Ali2017} argued that clusters containing $\sim$15\,PBHs should form by redshift $z\sim300$ (rather than $z\sim30$ as estimated by \citealt{Bird2016}), and therefore their density should be $1000\times$ higher (or cluster virial radius $10\times$ smaller) than in \citet{Bird2016}. 
These initial conditions change the approximate numerical values of the scalings derived in Section~\ref{sec:theory} to: $a_{\rm HS} \sim 10^5$\,au; $a_{\rm eject} \sim 10^4$\,au; $\tau_{\rm 3B} \sim 5$ Gyr $\ll \tau_{\rm 2B} \sim 5 \times 10^6$ Gyr.
Our numerical results can re-scaled by multiplying lengths by $\sim10$ and times by $\sim30$. Using these scalings, we find that the formation of the first hard binary occurs within $300$\,Myr, while the density drop that would significantly suppress two-body captures happens within $13.7$\,Gyr for all simulated clusters.

Finally, in Fig.~\ref{fig:4} we summarise the eccentricities and semi-major axes of hard binaries at the end of simulation: circles are binaries that are not bound to any other PBH, triangles and squares represent hard binaries respectively in triples and in multiple systems.
The typical outcome is the formation of binaries that are not bound to other PBHs, with occasional formation of triples, and only in a few cases the formation of multiple systems.
Binary semi-major axes ranges between $10^2 - 10^4\,$au with medians of $940, 700$ and $330\,$au (dashed vertical lines in the top panel of Fig.~\ref{fig:4}), respectively, for clusters (i), (ii) and (iii).
Only PBH binaries below and to the left of the black solid line, obtained by numerically integrating eq.~(5.14) of \citet{Peters1964} for $30 + 30\,$M$_\odot$ systems, can merge in $< 13.7\,$Gyr.
As all of the binaries lie significantly above the black line; none will merge due to GW emission within the Hubble time.
Even for the tightest binary with $a=100\,$au, $1-e$ needs to be as small as $10^{-4}$ (i.e. $e = 0.9999$) to merge in a Hubble time; however, the eccentricity would need to reach $1-e \approx 2\times10^{-6}$ once this binary is re-scaled to the \citet{Bird2016} cluster size.
High eccentricity can be induced by the Kozai-Lidov mechanism if the binary is in a hierarchical triple system.
We discuss this mechanism further in Section~\ref{sect:KL}.

\section{Discussion and Conclusions} \label{sec:discuss}

We carried out a set of simulations to investigate the dynamical evolution of clusters composed of 15 and 35 PBHs of 30\,M$_\odot$ and 10\,M$_\odot$ with an initial radius of $\sim1\,$pc. 
We find that in the considered regime hard binaries form via three-body interaction within 10\,Myr.
Subsequently binaries further harden via 2+1 interactions until they reach orbital separations of $\sim 10^2 - 10^3\,$au and eccentricities of $0-0.9999$ at $t=100\,$Myr.
Binaries with these orbital separation and eccentricities require more than a Hubble time to merge via GW radiation (cf. Fig.~\ref{fig:4}). 

Meanwhile, binaries act as a heat source in the cluster, driving its expansion and ultimate evaporation as energy is transferred from the binaries to the cluster through $2+1$ interactions. The expansion of the cluster lowers the rate of mergers through two-body captures by an order of magnitude or more by 14\,Myr (cf. Fig.~\ref{fig:3}), or by the age of the Universe if the cluster is re-scaled to the preferred density of \citet{Bird2016} and \citep{Ali2017}.

At the same time, the $2+1$ interaction rate drops for tight binaries and is further lowered as the density drops in expanding clusters (Eq.~\ref{eq:tau2+1}), ultimately becoming longer than the age of the Universe long before these binaries can merge through the emission of gravitational waves.  Consequently, binaries stall at the orbital separations represented in Fig.~\ref{fig:4}.
Below, we discuss two additional mechanisms that could enhance the dynamical merger rate, but find that these are unlikely to play a significant role for PBH clusters.

Therefore, we conclude that not accounting for cluster expansion through heating by binaries led \citet{Bird2016} to overestimate the PBH merger rate.  The actual merger rate of PBHs in such clusters is likely to be well below 1\,Gpc$^{-3}$yr$^{-1}$, so they do not contribute appreciably to the total binary black hole merger rate inferred from gravitational-wave observations.
\citet{Ali2017} pointed out that clusters composed of $\lesssim 3\times10^3$\,PBHs form earlier in the Universe at a higher density and should rapidly evaporate due to weak gravitational interactions. However, the scaling relations (equation~(99) of \citet{Ali2017}) are not necessarily valid in the small-$N$ regime considered in this paper. Our simulations show that clusters of 15 -- 35 PBHs do expand and evaporate because of the formation of hard binaries, and thus represent a useful numerical check of the approach of \citet{Ali2017}.

\subsection{Mergers induced by Kozai-Lidov mechanism} \label{sect:KL}

One rapid path to very high binary eccentricity, which reduces the GW merger timescale significantly, is through secular Kozai-Lidov (KL) oscillations \citep{Kozai,Lidov}.
This mechanism operates in hierarchical triples: two PBHs orbiting each other in a relatively tight inner binary plus a tertiary PBH orbiting the binary on a much wider outer orbit.
In this configuration the two orbits torque each other and exchange angular momentum, but not energy. Therefore, the orbits can change their eccentricities and relative inclination (typically on timescales much longer than their orbital periods), but not their semimajor axes.
\citet{Kozai}, \citet{Lidov} and subsequent work show that in Newtonian gravity between point masses, there is always a relative inclination of the outer to the inner orbit such that the inner binary can evolve to $e \sim 1$ from an arbitrarily small initial eccentricity through large-amplitude oscillations of the eccentricities and inclinations.
The time required to drive the inner binary from its minimum to maximum eccentricity is of the order of 
\begin{equation}
    \tau_{\rm KL} \sim \frac{16}{15} \frac{a_2^3}{a_1^{3/2}}\sqrt{\frac{M_1}{G m_3^2}} (1-e_2^2)^{3/2}
\end{equation}
where $a_1$ and $a_2$ are, respectively, the semimajor axes of the inner and outer binary, $M_1$ is the total mass of the inner binary, $m_3$ is the mass of the tertiary and $e_2$ is the eccentricity of the outer binary \citep[e.g.,][]{lid76}.  
For binaries formed in our simulation this timescale is smaller than a Hubble time for only $\sim 36$\% of the triples. When re-scaling to the reference clusters of \citet{Bird2016}, this time scale becomes $1000\times$ longer, so we can conclude that for most triples, KL oscillations are inefficient at driving up inner binary eccentricities. 

However, it is challenging to definitively rule out a contribution of KL oscillations to the PBH merger rate. Triples will form generically as a result of $2+2$ interactions, with the expected ratio of such interactions to $2+1$ interactions of order the ratio of the number of binaries to the number of single stars in the cluster, or $\sim 0.1$ for our typical clusters.  The presence of triples in the simulated clusters is illustrated in Fig.~\ref{fig:4}. 
Only one of the triples formed in the 30 simulated clusters shown in
Fig.~\ref{fig:4} has a KL timescale below the age of Universe after re-scaling. This triple has re-scaled inner and outer binary semimajor axes of $4.2\times 10^4$\,au and $2.8 \times 10^6$\,au, respectively; its inner binary eccentricity would have to grow to a very large eccentricity $1-e\approx 4 \times 10^{-7}$ in order for it to merge through gravitational-wave emission in the age of the Universe.
It is therefore possible that hierarchical triples could form and be driven to sufficiently high eccentricities by KL oscillations to merge in a small fraction of PBH clusters.

\subsection{Mergers due to dynamical inspirals}

In a cluster containing binaries, 2+1 interactions can be efficient in dynamically forming highly eccentric inspiraling binaries. 
A significant fraction of the binary-single encounters result in resonant interactions, in which the three PBHs wander for a long time on chaotic orbits and approach each other repeatedly \citep{heggie75}. 
In particular, during these chaotic encounters two PBHs can pass sufficiently close to capture through the emission of GW radiation and even merge while the system is still in resonance.  This last outcome is rare.  However, the cross section of binaries is larger than that of a single PBH for two-body capture.  In particular, the cross section of 2+1 interactions is larger for binaries with larger orbital separations, and the overall cross-section for dynamical inspirals scales as $a^{2/7}$ with the target binary semi-major axis for equal-mass binaries \citep[e.g.,][]{gul06,Samsing2014,sam19}. 
Thus, wide binaries formed in our simulations can potentially open another merger channel.

The cross section of these encounters can be estimated using eq.~(36) from \citet{Samsing2014}:
\begin{equation}
    \sigma_{\rm insp} 
    \simeq 5 \times 10^{-8} \left( \frac{a}{10^3\text{au}}\right)^{2/7} \left( \frac{m}{30{\text \Msun}}\right)^{12/7}\left( \frac{v_{\rm disp}}{1.5 \text{km\,s}^{-1}}\right)^{-2} \quad \text{pc}^2. 
\end{equation}
We can compare this cross section to the cross section for direct two-body capture given in Eq.~\eqref{eqn:2brate}, using the \citet{Bird2016} cluster parameters: $m = 30\,{\text \Msun}$,  $v_\textrm{disp} = 0.15$\,km s$^{-1}$, $a=a_\textrm{HS}\sim 10^6$\,au (the widest stable binaries provide the greatest contribution to dynamical interactions).  With these values, $\sigma_{2\textrm{B}} \approx 2 \times 10^{-6}$ pc$^2$, while $\sigma_\mathrm{insp} \approx 4 \times 10^{-5}$ pc$^2$, a factor of 20 greater.  However, the fraction and hence number density of binaries is only $\sim 0.1$ of that of single stars for typical simulated clusters (see Fig.~\ref{fig:2}), reducing the relative contribution for this channel by a factor of $\sim 10$.  Consequently, the overall rate of captures during three-body interactions is comparable to the rate of direct two-body captures for cluster parameters of interest.  Both rates will drop as $n v_{\rm disp}^{-18/7} \sim n a^{2/7} v_{\rm disp}^{-2} \sim R^{-12/7}$ as the cluster expands.

\vspace{0.35in} 

In this work we re-examined the merger rate of PBH binaries formed in the subdominant late Universe channel by accounting for the first time for the full spectrum of dynamical interactions, including binary formation and subsequent cluster heating and expansion.
We find that the contribution of $\lesssim 10^3\,$M$_\odot$ PBH clusters to the binary black hole merger rate (if no initial binaries are present in the cluster) falls by an order of magnitude or more, to well below 1\,Gpc$^{-3}$ yr$^{-1}$.   
If PBHs comprise only a fraction $f_{\rm pbh}$ of the DM, the event rate will scale as $ f_{\rm pbh}^{53/21}$ \citep{Bird2016}.
Gravitational captures would thus make a minimal contribution to the observed rate of black hole mergers, making detections of merging binary black holes formed through this process unlikely. Consequently, if PBHs are formed in such low-mass clusters but not in tight binaries, gravitational-wave observations cannot observe PBHs or constrain their contribution to the dark matter content of the Universe.

\section*{Acknowledgements}
We acknowledge the Kavli Foundation and the DNRF for supporting the 2017 Kavli Summer Program, and thank the Niels Bohr Institute for its hospitality while part of this work was completed. 
V.K.~acknowledges support from the Netherlands Research Council NWO, specifically WRAP Program (grant 648.003004 APP-GW) and the Rubicon Program (grant 019.183EN.015).
IM and MCM acknowledge support from the Munich Institute for Astro- and Particle Physics (MIAPP), which is funded by the Deutsche Forschungsgemeinschaft under Germany's Excellence Strategy EXC-2094-390783311.  IM is a recipient of the Australian Research Council Future Fellowship FT190100574. MCM was also supported by a Visiting Researcher position at Perimeter Institute for Theoretical Physics, and by the Radboud Excellence Initiative for supporting his stay at Radboud University, in the last stages of this project.\\ Simulations in this paper made use of the REBOUND code which can be downloaded freely at http://github.com/hannorein/rebound.
\addcontentsline{toc}{section}{Acknowledgements}

\bibliographystyle{mnras}
\bibliography{biblio}

\bsp	
\label{lastpage}
\end{document}